\documentclass[aps,twocolumn,prl,showpacs]{revtex4}
\usepackage{epsfig}
\usepackage{graphicx}
\usepackage{amsmath}

\begin{document}

\title{ Resonating valence bond  wave function for the two dimensional 
fractional spin liquid}

\author{S.~Yunoki}
\author{S.~Sorella}
\affiliation{ INFM-Democritos, National Simulation Centre, 
 and SISSA, I-34014 Trieste, Italy}
\date{\today}
\begin{abstract}

The unconventional 
low-lying spin excitations,  recently observed in neutron scattering 
experiments on ${\rm Cs_2 Cu Cl_4}$, are explained with a spin liquid 
 wave function. 
The dispersion relation  as well as the wave vector 
of the incommensurate spin correlations  are well reproduced within 
a projected BCS wave function with gapless and fractionalized spin-1/2 
excitations around the nodes of the BCS gap function. 
The proposed wave function is shown to be very accurate for 
one-dimensional spin-1/2 
systems, and remains similarly accurate in the two-dimensional model 
corresponding to 
${\rm Cs_2 Cu Cl_4}$, thus representing a good ansatz for 
describing spin fractionalization in two dimensions.
\end{abstract}

%%%%%%%%%%%%%%%%%%
\maketitle

The fractionalization of spin excitations in one-dimensional (1D) 
spin-1/2 antiferromagnetic (AF) Heisenberg systems was conjectured by 
Faddeev and Takhtajan~\cite{faddeev} more than 20 years ago, and by now 
the concept of ''spinons'', as spin-1/2 elementary excitations for 1D quantum 
AF systems, has been well established both theoretically~\cite{bethe} and 
experimentally~\cite{exp}. 

An important and intriguing issue raised by Anderson~\cite{anderson} 
after the discovery of 
high-$T_{\rm C}$ superconductors is whether 
fractionalized gapless spinons can be defined even in higher dimensions. 
Many theoretical studies have been done so far analytically~\cite{millis} 
and numerically~\cite{sandvick}, mainly by considering weakly coupled chains. 
Most studies suggest that unconventional 1D features are very unlikely to 
occur in higher dimensions since more conventional states, {\it e.g.}, 
magnetically ordered, are stabilized 
with a weak inter-chain coupling. 

\begin{figure}[hbt]
\includegraphics[width=3.3cm,angle=0]{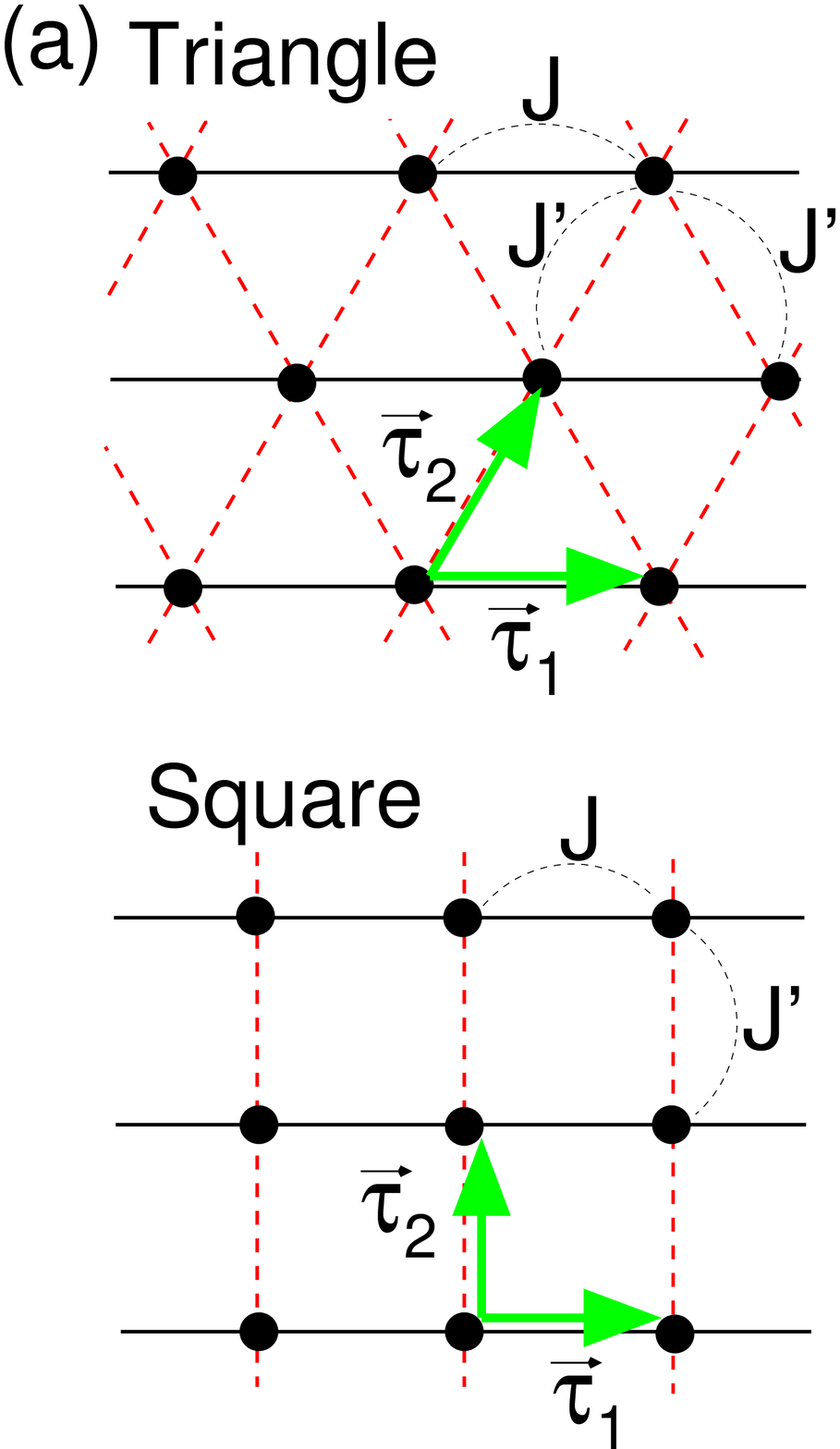}
\includegraphics[width=3.9cm,angle=-0]{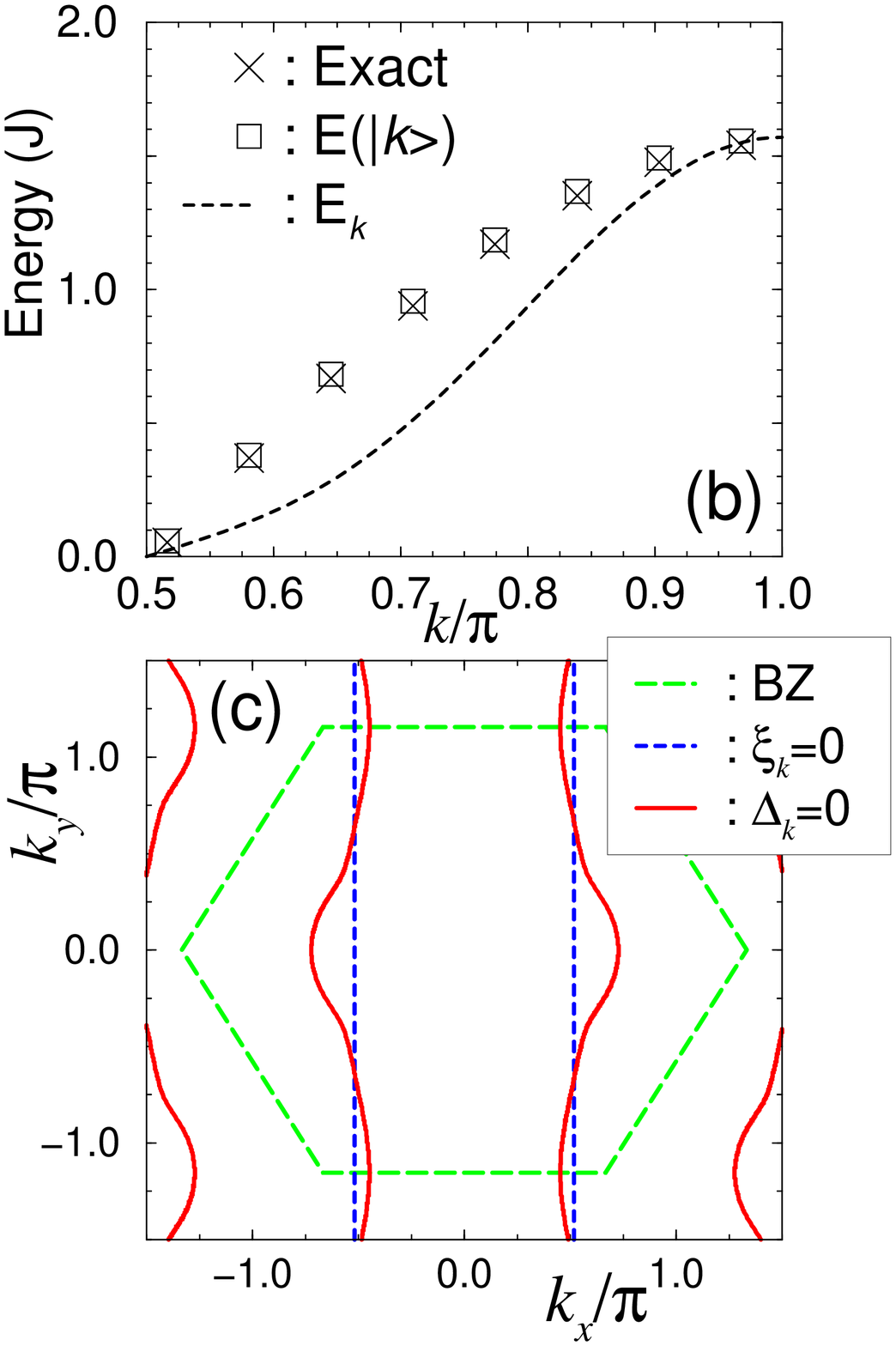}
\begin{center}
\caption{
(a) Lattice geometries studied. 
${\vec\tau}_1$ and ${\vec\tau}_2$ are the primitive translational vectors.
(b) Lowest energy spin-$1/2$ excitations of the 1DAFHM as a function 
of momentum ${\bf k}$. Dashed line is the BCS spectrum $E_{\bf k}$ scaled 
by a factor to match the exact bandwidth. 
(c) Loci of ${\bf k}$-points where $\xi_{\bf k}=0$ (dashed lines) and 
$\Delta_{\bf k}=0$ (solid lines). The boundary of the first Brillouin 
zone (BZ) for the triangular lattice is also denoted by long dashed lines. 
}
\label{lattice}
\end{center}
\end{figure}

The same question can be also addressed experimentally. 
A series of recent inelastic neutron scattering experiments on 
${\rm Cs_2 Cu Cl_4}$ by Coldea {\it et al}~\cite{coldea,coldealong} 
showed that, as in 1D systems~\cite{exp}, the spectrum 
on this material consists of a broad 
incoherent continuum at each momentum, interpreted as spin fractionalization, 
despite the fact that the system is clearly 
two-dimensional (2D). 
It was also found that 
the system is described  by the following 2D spin-1/2 AF Heisenberg 
model (2DAFHM) on the triangular lattice (see Fig.~\ref{lattice}(a)): 
\begin{equation} \label{model}
H= J \sum\limits_{\langle {\bf i},{\bf j}\rangle}  
\vec S_{\bf i} \cdot \vec S_{\bf j}   + J^\prime 
 \sum_{\langle\langle {\bf i},{\bf j}\rangle\rangle}  
\vec S_{\bf i} \cdot \vec S_{\bf j}
\end{equation}
with the intra-chain coupling $J=0.374$ meV and the inter-chain 
coupling $J^\prime/J=0.33$~\cite{coldea}. Here 
the symbol  $\langle {\bf i},{\bf j}\rangle$ 
($\langle\langle {\bf i},{\bf j}\rangle\rangle$) indicates nearest neighbor 
sites along 
the chain (between different chains). 
Additional terms such as the Dzyaloshinskii-Moriya interaction and the 
coupling in the third direction,  both of the same order 
$\simeq 0.02~{\rm meV}$~\cite{coldealong}, 
are certainly relevant to explain the various low-temperature 
phases~\cite{coldea}. 
However in this study we ignore these much smaller terms as 
our main purpose is to determine the minimal model and its ground state 
wave function (WF) which may lead to spin fractionalization in the 
2D system.

 In this Letter  we  propose a projected BCS WF for the ground state 
of Eq.~(\ref{model}): 
$|\Phi\rangle=\hat{\cal P}|{\rm BCS}\rangle$, 
where $|{\rm BCS}\rangle$ is the ground state of the following mean-field BCS 
Hamiltonian, 
\begin{equation} \label{bcsham}
H_{\rm BCS} = \sum_{{\bf k},s} \xi_{\bf k}  
c^{\dag}_{{\bf k},s} c_{{\bf k},s}
+ \sum_{\bf k} \left[ \Delta_{\bf k} 
c^{\dag}_{{\bf k},\uparrow} c^{\dag}_{-{\bf k},\downarrow} 
+ {\rm h.c.} \right]. 
\end{equation} 
Here $\xi_{\bf k}=\epsilon_{\bf k}-\mu$, 
$\epsilon_{\bf k}=-2\cos({\bf k}\cdot{\vec \tau}_1)$, 
$\mu$ is the chemical potential, 
$c^{\dag}_{{\bf k},s}$ the  creation operator of an electron with 
momentum ${\bf k}$ and spin $s=\pm1/2$, 
and $\Delta_{\bf k}$ the real gap function 
for singlet pairing with ${\rm A_1}$ symmetry. 
The  variational state $|\Phi\rangle$ for the spin Hamiltonian $H$ 
is obtained  by applying to $|{\rm BCS}\rangle$ the Gutzwiller 
projector $\hat{\cal P}$ 
onto the subspace of singly occupied sites. 
The quantities  $\Delta_{\bf k}$ and $\mu$ are determined following the 
variational principle, {\it i.e.}, by minimizing the energy 
$E(|\Phi\rangle)=\langle\Phi|H|\Phi\rangle/\langle\Phi|\Phi\rangle$. 
For this purpose, the Fourier transform $\Delta_{{\bf i},{\bf j}}$ of 
$\Delta_{\bf k}$ is truncated up to the third distance 
$|{\bf i}-{\bf j}|$ along the 
chain so that 10 independent parameters for $\Delta_{\bf k}$ are 
optimized along with $\mu$~\cite{sorella,note}. 
Note that whenever $\mu$ is finite, $H_{\rm BCS}$ no longer 
possesses 
particle-hole symmetry and therefore $|\Phi\rangle$ violates the Marshall 
sign rule~\cite{sorellanew}. 
In the following, coupled $L$-site chains for a total number 
$N=L^2$ of sites with periodic boundary conditions are used unless 
otherwise stated. In order to explore the quality of $|\Phi \rangle$, 
the variance of the energy, 
$\sigma^2(|\Phi\rangle )={\langle\Phi|(H-E(|\Phi\rangle))^2|\Phi\rangle/
\langle \Phi | \Phi \rangle }$, is also calculated~\cite{note2}. 
For comparison, the same spin liquid WF $|\Phi\rangle$  
is applied to other models, the uncoupled Heisenberg chains  
with $J^\prime=0$ and the 2DAFHM on the square and the triangular lattices 
with $J^\prime/J=0.33$ and $J'/J=1$, respectively 
(see Fig.~\ref{lattice}(a)). 
For each model the gap function $\Delta_{\bf k}$ 
and $\mu$ are optimized independently. 
 
The main point of this Letter is that  the variational approach represents an 
accurate theoretical tool for extending the notion of spinons in higher 
dimensions. With this approach we are able to reproduce the qualitative, as 
well as the quantitative  features of the experiments, confirming the 
possibility of  spin fractionalization in 2D with a transparent theoretical 
framework. Following Wen~\cite{wenprb}, we also  show  that the excitations 
of $H_{\rm BCS}$ are related to the physical excitations of the spin 
Hamiltonian.

As a first step, let us verify that the highly non trivial low-energy 
spectrum of the 1D spin-1/2 AF Heisenberg model (1DAFHM) 
with nearest neighbor coupling $J$ is well reproduced 
by the WF $|\Phi\rangle$. 
It has been well known that $|\Phi\rangle$ describes 
almost exactly the ground state WF~\cite{superb,gros}. 
Here we show that even the low-lying excited states can be constructed 
from $|\Phi\rangle$. 
For the 1D model, the optimized $\mu$ is zero and therefore the BCS spectrum  
$E_{\bf k}=\sqrt{ \xi_{\bf k}^2+\Delta_{\bf k}^2}$ 
of Eq.~(\ref{bcsham}) has gapless excitations at  ${\bf k}=\pm \pi/2$, 
exactly as the spinon spectrum of the 1DAFHM does~\cite{bethe}. 
Since the elementary excitations of $H_{\rm BCS}$  with energy $E_{\bf k}$
 are described  by the Bogoliubov modes, $\gamma^{\dag}_{{\bf k},s}$, 
the simplest variational state for the spinon at momentum ${\bf k}$ is
$|{\bf k}\rangle= \hat{\cal{P}} \gamma_{k,\downarrow}^\dag |{\rm BCS}\rangle$. 
To see whether this state $|{\bf k}\rangle$ corresponds to 
a spinon state, we consider a ring with $odd$ number of sites $L=31$ 
and {\it z}-component of the total spin $S_z^{\rm tot}=-1/2$. 
For this case it is known that a well defined  spinon exists only for half 
of the total Brillouin zone ($\pi/2 \le |{\bf k}| \le \pi $). 
As shown in Fig.~\ref{lattice}(b), for this branch, the WF 
$|{\bf k}\rangle$ represents  fairly well the excited state with a spinon
at  momentum ${\bf k}$,  
as can be verified by the good accuracy in energy. 
Notice that, although the projection $\hat{\cal P}$ 
is crucial to gain a quantitative agreement for the spectrum, 
the  BCS  spectrum $E_{\bf k}$ already  gives  
a qualitatively correct feature of gapless  
 excitations  with finite spinon velocity at the 
right momentum ${\bf k}=\pi/2$ (see Fig.\ref{lattice}b). 
It is also possible to obtain an accurate description for the 
remaining branch of 
the spectrum by using a similar variational 
WF with three Bogoliubov modes.

\begin{figure}[hbt]
\includegraphics[width=4.7cm,angle=-90]{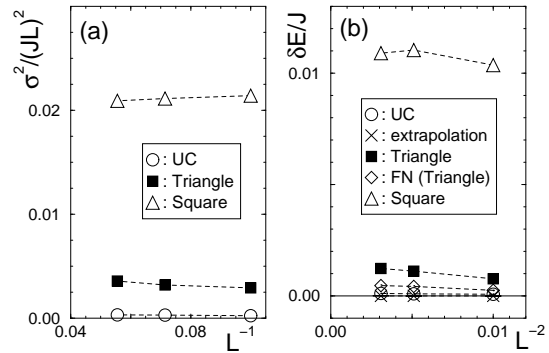}
\begin{center}
\caption{
(a) Variance of energy  $\sigma^2 (|\Phi\rangle)$ 
 for $N=L\times L$ clusters. UC stands 
for uncoupled chains.  
(b) Estimate of the variational error in the energy per site. The  
 FN variational error for the anisotropic triangular lattice and 
the variance extrapolated error for uncoupled chains are also plotted.   
}
\label{variance}
\end{center}
\end{figure}

Even in 2D, the elementary excitations of $H_{\rm BCS}$ define quite 
naturally, after Gutzwiller projection, the spin-1/2 fractionalized 
excitations of the 2DAFHM~\cite{wenprb}. In order to consider the 
possibility of spin fractionalization in 2D, 
it is therefore important to show that the WF $|\Phi\rangle$ 
can appropriately describe the ground state of Eq.~(\ref{model}). 
The most important effect of frustration 
on the proposed WF $|\Phi\rangle$ is generally to change $\mu$ to 
a nonzero value. 
This should be contrasted to the case of the 2DAFHM on the square lattice with 
no frustration such as the one shown in Fig.~\ref{lattice}(a), where 
the optimized $\mu$ is always zero. 
Therefore the Marshall sign rule for the state $|\Phi\rangle$ is violated 
on the triangular lattice, and the gapless excitations of $H_{\rm BCS}$ 
are no longer located at commensurate 
points. For example, the optimized value  for $N=18\times18$ is 
$\mu =0.110(4)$. Thus one can make a spin-1 gapless excitation at wave vector 
$ Q_x= 2 \cos^{-1} ( -\mu/2 )\sim1.035(2)\pi$, which is   
close to the experimentally observed incommensurate wave vector on 
${\rm Cs_2 Cu Cl_4}$~\cite{coldea}. Indeed, as shown in Fig.~\ref{lattice}(c), 
the nodal lines of $\Delta_{\bf k}$ intersect the lines defined by 
$\xi_{\bf k}=0$, and thus four gapless modes exist.

To check the quality of the proposed WF $|\Phi\rangle$, 
the variance $\sigma^2(|\Phi\rangle)$ is calculated in 
Fig.~\ref{variance}(a), where for comparison the variances are also 
presented for uncoupled chains with $J^\prime=0$ (spin liquid) and for 
the 2DAFHM on the square lattice with $J^\prime/J=0.33$ (non spin liquid).  
As seen in Fig.~\ref{variance}(a),  
the variance for the triangular case is much smaller than the variance 
corresponding to the square lattice and instead this 
value is much closer to the one for the uncoupled chains.  
To further explore the quality of the WF, we also calculate in 
Fig.~\ref{variance}(b) the error of the variational energy from the exact 
value $e_0$: 
$\delta E=E(|\Phi\rangle)/N- e_0$. 
Since the exact energy for the triangular lattice case is not known, 
we estimate $e_0$ by extrapolating to zero variance 
the variational energies corresponding to  
$|\Phi\rangle$ and to the one Lanczos step WF 
$|\Phi_{\rm 1LS}\rangle = ( 1+\alpha H )|\Phi\rangle$ 
with optimized $\alpha$~\cite{sorella}. 
Indeed this procedure works well for the uncoupled chains  as shown in 
Fig.~\ref{variance}(b). Though the WF  $|\Phi\rangle$ 
for the triangular lattice is not as accurate as for the uncoupled chains, 
it remains of high quality, much better than the square lattice case. 
It should be also emphasized that the corresponding values for the best 
available variational WF with explicit AF order of the widely studied 
isotropic 2DAFHM on the square 
lattice with $J'=J$~\cite{ogata} 
are $\sigma^2\sim0.015$ and $\delta E\sim0.004$ in the same unit as in 
Fig.~\ref{variance}. 
Also comparing these values to ours, the quality of the 
WF $|\Phi\rangle$ for $H$ is exceptionally 
good for 2DAFHM's.

Though the variational approach can  provide accurate upper bounds for 
 the energy, it may suffer the well known difficulty that 
the low energy modes are usually insensitive to the total energy, and 
sometimes  completely different WF's give more or less the same 
energy~\cite{hei}. 
In order to reduce this difficulty,  quantum Monte Carlo methods,  such 
as a recently developed fixed node (FN) technique, may 
be used~\cite{sorella,ceperley}. 
Within this more powerful FN approach, the variational WF $|\Phi\rangle$ 
is used to approximate only the signs of the ground state WF. 
In the basis of electron configuration $|x\rangle$ all the off-diagonal 
matrix elements of the Hamiltonian such that 
$s_{x^\prime x} = \Phi (x^\prime)  \langle x^\prime| H | x \rangle 
 \Phi(x)  < 0$ are considered exactly. 
Instead all the ones with $s_{x^\prime x} > 0$  
are treated semiclassically and traced to the diagonal term, defining 
an effective FN Hamiltonian $H_{\rm FN}$. 
Since the semiclassical approximation is usually good at low energy, 
we believe that this approach provides a sensible test to examine the 
stability of the variational WF at low energy. 
Indeed, if, for a properly chosen WF, $s_{x^\prime x}$ is always non 
positive for $x\ne x'$, 
the FN approximation is exact and may change completely the low energy 
properties of the variational WF. 
Namely, for the 2DAFHM on the 
square lattice, it is possible to obtain the exact AF ordered state 
provided that the variational WF satisfies the Marshell sign rule, 
a condition that does not necessarily imply AF order.  
In the frustrated case no WF's which fulfill $s_{x^\prime x}<0$ 
for $x\ne x'$ are known. 
Therefore, in order to verify the quality of the FN approximation, we 
estimate by quantum Monte Carlo method the fraction $w$ between the 
positive and the negative off-diagonal elements. 
It is found that, due to the quality of our WF, the ratio $w$ remains 
very small ($\sim0.03$), implying that $H_{\rm FN}$ contains a large 
fraction ($\simeq 1-w$) of matrix elements equal to the exact ones, 
showing that, in the present case, 
$H_{\rm FN}$ represents a reliable approximation of $H$. 
%%%%%
%%%%%
We finally remark that  
the ground state $|\Psi_0\rangle$ of $H_{\rm FN}$ can be used 
as a variational state of $H$ with lower energy~\cite{ceperley},  
$E(|\Psi_0\rangle)\le E(|\Phi\rangle)$, as clearly seen in 
Fig.~\ref{variance}(b).

\begin{figure}
\includegraphics[width=5.5cm,angle=-90]{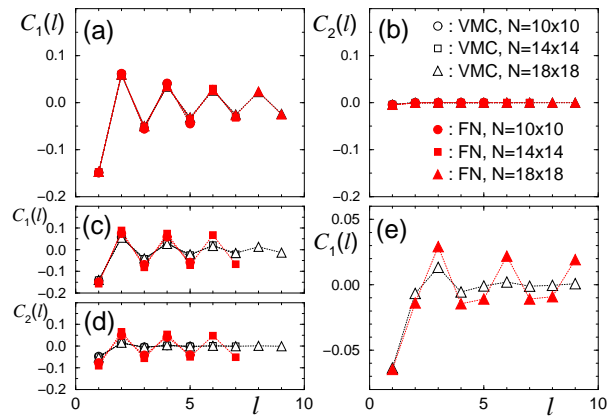}
\begin{center}
\caption{
Spin correlation functions 
$C_{i}(l)=\langle S_z({\vec r}) S_z({{\vec r}+l{\vec\tau_i}})\rangle$ 
as a function of distance $l$ along the $\vec\tau_i$ directions for the 
2DAFHM on the triangular lattice with $J'/J=0.33$ (a,b) and $J'/J=1.0$ (e), 
and on the square lattice with $J'/J=0.33$ (c,d). 
Open and solid marks are for the variational (VMC) and FN calculations, 
respectively. 
Lines are guides to the eye.   
}
\label{SzSz}
\end{center}
\end{figure}

 In Fig.~\ref{SzSz}, the spin correlation functions (z-component) 
$\langle\psi|S_z({\bf i})S_z({\bf j})|\psi\rangle/\langle\psi|\psi\rangle$
are calculated for the 2DAFHM on both the triangular and square lattices, 
using the variational ($|\psi\rangle=|\Phi\rangle$) and the 
FN ($|\psi\rangle=|\Psi_0\rangle$) methods. 
For the square geometry, it is clear that  
the FN approach considerably enhances the large distance correlations, 
suggesting that the spin liquid state is unstable toward AF long-range order, 
in agreement with previous studies~\cite{sandvick}. 
A completely  different scenario  is evident  for the triangular geometry  
where these long distance correlations 
do not appreciably change with respect to the variational 
$|\Phi\rangle$, strongly indicating that the spin liquid state $|\Phi\rangle$ 
 is stable, namely close to the exact ground state, at least 
 for  the clusters studied.

In order to show further the reliability of our calculations, 
the spin correlation functions for the isotropic 2DAFHM on the triangular 
lattice with $J'=J$ are also calculated in Fig.~\ref{SzSz}(e). 
Here a $\hat{\cal P}|{\rm BCS}\rangle$ with good variational energy 
is obtained with $\xi_{\bf k}=0$, and 
a simple gap function with $\Delta_{{\bf i},{\bf j}}=\pm 1$ 
restricted up to second nearest distances and with properly chosen 
signs~\cite{sondhi}. Notice that, even by using such a simple spin liquid 
WF with a rather large value of $w\sim22\%$, the FN method correctly 
reproduces the classical three sublattice antiferromagnetic 
correlations~\cite{af}.

\begin{figure}
\includegraphics[width=5.25cm,angle=-90]{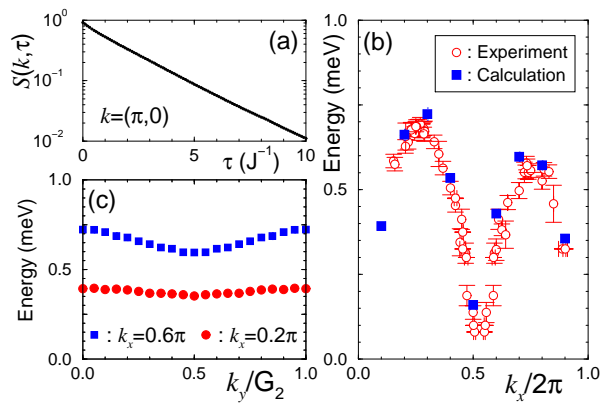}
\begin{center}
\caption{
(a) A typical semi-log plot of the imaginary time correlation function. 
(b) Calculated spin-1 excitation energies (solid squares) 
along ${\bf k}//{\vec\tau}_1$ for the anisotropic triangular lattice with  
$N=10\times20$, $J'/J=0.33$, and $J=0.374$ meV. 
(c) The same as in (b) but along ${\bf k}\perp{\vec\tau}_1$ for different 
$k_x$. Here $G_2=4\pi/\sqrt{3}$. For comparison, 
experimental data on ${\rm Cs_2 Cu Cl_4}$~\cite{coldealong} (open circles) 
are also shown in (b) 
}
\label{sq}
\end{center}
\end{figure}

Encouraged by the above results, we now consider the low-lying excitations 
on the spin liquid state $|\Phi\rangle$ for the 2DAFHM on the triangular 
lattice with $J'/J=0.33$. To this end the FN method is applied.  
Since $w$ is rather small for this case, we expect that 
this approach should give a reliable description of the exact 
excitation spectrum of $H$.
By using the forward walking technique~\cite{calandra}, one can  
evaluate the imaginary time ($\tau$) evolution of the following quantity: 
\begin{equation} \label{sqimag}
S({\bf k},\tau) =
 { \langle\Phi |S_z({\bf k}){\rm e}^{-\tau H_{\rm FN}}S_z(-{\bf k})|\Psi_0\rangle  \over
  \langle\Phi | {\rm e}^{-\tau H_{\rm FN}} |\Psi_0  \rangle  }, 
\end{equation}
where $S_z({\bf k}) = \sum_{\bf r} {\rm e}^{i{\bf k}\cdot{\bf r}}
S_z({\bf r})/\sqrt{N}$. 
By simple inspection, the lowest spin-1 excitation energy $E^{S=1}_{\bf k}$ of
$H_{\rm FN}$  can be obtained by fitting the large $\tau$  behavior of
$S({\bf k},\tau) \propto {\rm e}^{ - E^{S=1}_{\bf k} \tau }.$  
An example of how the method works is given in Fig.~\ref{sq}(a). 
For large $\tau$, $\ln S({\bf k},\tau)$ is almost linear and thus 
we can estimate 
the lowest excitation energy for each ${\bf k}$. The results are 
summarized in Fig.~\ref{sq}(b) and (c). 
While the spectrum has a strong 
1D characteristic, it shows a visible and non trivial dispersion 
in the ${\bf k}$-direction perpendicular to $\vec\tau_1$. 
Moreover, as seen in Fig.~\ref{sq}(a), the linear fit is not perfect in 
the intermediate times $\tau J\simeq 1$ implying that there are higher energy 
($\sim J$) contributions to the spectrum, suggesting that the full spectrum 
of the dynamical structure factor may be consistent with the incoherent 
continuum observed in the experiments. 
The extremely good agreement between the calculated spectrum and the 
available experimental data~\cite{coldea}, shown in Fig.~\ref{sq}(b), 
indicates that the 2D model 
of Eq.~(\ref{model}) describes correctly the experiments, thus 
supporting the appearance of a spin liquid state. 
 Since  no sizable  magnetic order was found in this 2D  model 
(Fig.~\ref{SzSz}), it is very likely  that  
the experimentally 
observed ordered phases at very low temperatures should  be due only to 
3D effects~\cite{coldealong}. 
We cannot exclude the possibility that a small spin gap could remain  in the 
thermodynamic limit, as our numerical 
resolution is limited in the  available  finite size clusters.

In conclusion, we have shown that, by studying several models, 
a simple variational approach is capable of describing the ground state 
properties as well as the low-lying excitations of a spin liquid not only 
for 1D systems but also for a 2D frustrated spin-1/2 model. 
Within this approach it appears that, when the  $H_{\rm BCS}$  spectrum 
remains gapless at commensurate ${\bf k}$ as in a non frustrated square 
lattice, the spin fractionalized state undergoes a magnetic instability. 
By contrast, in the frustrated triangular case
the $H_{\rm BCS}$ spectrum becomes gapless at incommensurate ${\bf k}$ 
as soon as $J'/J>0$, and it appears much more 
difficult to destabilize the fractionalized state by a magnetic phase 
transition.    
It should be emphasized that both the experimental and the present numerical 
work consistently suggest that a spin liquid with gapless spin-1/2 
excitations appears possible in 2D, its stability being intimately related 
to the incommensurate momenta  of the gapless  modes.

Our results may have some impact even for high-$T_{\rm C}$ superconductors, 
where the carrier doping 
naturally leads to incommensurate nodal Fermi points. 
The existence of a nodal spin liquid has been also conjectured  
before for interacting Bose systems~\cite{doniach,fisher} 
where a Bose metal is the natural bosonic counterpart of a  spin liquid  
with gapless spin excitations and no magnetic order.

We acknowledge R. Coldea for sending us the experimental data and F. Becca 
for providing us the exact diagonalization data.
This work is partially supported by INFM-MALODI and MIUR-COFIN 2001.

%%%%%%%%%%%%%%%%%%%%%%%%%%%%%%%%%%%%%%
%%%%%%%%%%%%%%%%%%%%%%%%%%%%%%%%%%%%%%


\begin{thebibliography}{99}
\bibitem{note} The nearest-neighbor hoppings  along ${\vec\tau_2}$ and 
$\vec\tau_2-\vec\tau_1$ were also optimized and found 
to be negligibly small.
\bibitem{note2} Only when $|\Phi\rangle$ is an exact eigenstate, 
$\sigma^2( |\Phi\rangle)=0$, and the smaller is its value, the better is 
the quality of $|\Phi\rangle$.
\end{thebibliography}
\end{document}